\documentclass[twoside,a4paper]{bjp}
\usepackage{epsfig,cite}
\usepackage{graphics}
\usepackage{amssymb,amsmath}
\usepackage{times}
\begin{document}

\sloppy \raggedbottom

 \setcounter{page}{1}



\title{Shapes and Dynamics from the Time-Dependent Mean Field}

\runningheads{Shapes and Dynamics from TDHF}{P. D. Stevenson, P. M. Goddard and A. Rios}

\begin{start}
\author{P.~D.~Stevenson}{}, \coauthor{P.~M.~Goddard}{},
\coauthor{A.~Rios}{}

\address{Department of Physics, University of Surrey, Guildford, GU2 7XH, Surrey, United Kingdom}{}


\begin{Abstract}
Explaining observed properties in terms of underlying shape degrees of freedom is a well--established prism with which to understand atomic nuclei.  Self--consistent mean--field models provide one tool to understand nuclear shapes, and their link to other nuclear properties and observables.  We present examples of how the time--dependent extension of the mean--field approach can be used in particular to shed light on nuclear shape properties, particularly looking at the giant resonances built on deformed nuclear ground states, and at dynamics in highly-deformed fission isomers.  Example calculations are shown of $^{28}$Si in the first case, and $^{240}$Pu in the latter case.
\end{Abstract}

\end{start}

\section[]{Introduction}
The fact that atomic nuclei can take on wide variety of shapes in different intrinsic states is evidenced by the appearance of excited states identified within collective models as e.g. rotational bands, vibrational states, gamma bands \cite{Cas01}, or within microscopic approaches such as GCM in terms of wave functions peaked at particular deformations \cite{Ben04}.  While deformation is not an observable \textit{per se}, the link between deformation and observable quantities is a well-established cornerstone of the interpretation of nuclear spectra.

A self-consistent mean--field based approach using Skyrme--Hartree--Fock \cite{Ben03,Sto07,Ben03b} is one starting point for the study of nuclear shapes \cite{Ste03,Ste05}.  Of the many kinds of extension beyond the static mean field that can probe nuclear shapes, we concentrate here on results from time-dependent Hatree--Fock calculations.  The time--dependent Hartree--Fock (TDHF) method \cite{Sim12} uses an Ansatz that the nuclear wave function is represented at all times by single Slater Determinant (often with occupation probabilities smeared out by pairing interactions), which evolves as time progresses.  In some sense the approach remains a fully mean--field approach, though since there are many Slater Determinants involved when one considers all times in a calculation, there is also a sense in which the TDHF approach is a beyond--mean--field one.  Certainly, in the treatment of small-amplitude vibrations of the giant resonance type, one typically takes the Fourier transform in the time coordinate of a time series of and observable extracted from the set of Slater Determinants, and one gains results formally equivalent to the Random Phase Approximation \cite{Rin80}.

In this contribution, we give a summary of the formalism of TDHF and the Skyrme interaction, and present results for some cases in which one can learn about nuclear shapes from the dynamics which result from TDHF calculations.  

\section{Time--Dependent Hartree--Fock}
The time--dependent Hartree--Fock equations were originally posited by Dirac in the 1930s \cite{Dir30}.  Calculations in the case of nuclear physics came only after computational power rendered them practical, in the 1970s and 1980s \cite{Neg82,Dav85}.  Since then, realizations of the TDHF equations have been made with few or no restrictions on the symmetry of the system being considered \cite{Kim97,Mar14} -- meaning for example that any nuclear shape can be studied -- and also with quite sophisticated forms of the effective interaction, providing one sticks to those of the Skyrme type \cite{Fra12} (but including many or all the ways in which they have been extended).

The detailed formalism can be found elsewhere \cite{Sim12,Mar14}, but the short description is that one starts from a static Hartree--Fock solution of some sort, in which the density is found self-consistently together with a density--dependent Hartree--Fock Hamiltonian such that
\begin{equation}
  [h_\mathrm{HF}(\rho),\rho]=0.
\end{equation}
In time--dependent Hartree--Fock the equation now reads
\begin{equation}
  [h_\mathrm{HF}(\rho),\rho]=i\hbar\partial_t\rho.
\end{equation}
In order for this to give a non--trivial time--evolution to a density which describes an eigenstate of the Hartree--Fock Hamiltonian, an external perturbation is usually provided in some form.  The most common examples are; (a) that two nuclei are placed together on a grid and are perturbed with initial velocity vectors setting them on a collision course to study the resulting reaction (see e.g. \cite{Uma08,Uma10,Dai14,Ste15} for selected recent examples); (b) initialization of a single nucleus with a shaped field (usually, but not always, multipolar) to instigate a giant resonance (e.g. \cite{Nak05,Ste04,God13}); (c) giving an effective perturbation by not starting from an eigenstate of the HF Hamiltonian, but instead some physically meaningful state connected to the true Hamiltonian -- e.g. a multiquasiparticle state, or a solution of the static HF equations subject to some constraint, such as the quadrupole deformation.

\section{Giant Resonances as Probes of Shape}
\begin{figure}[tb!]
  \includegraphics[width=\textwidth]{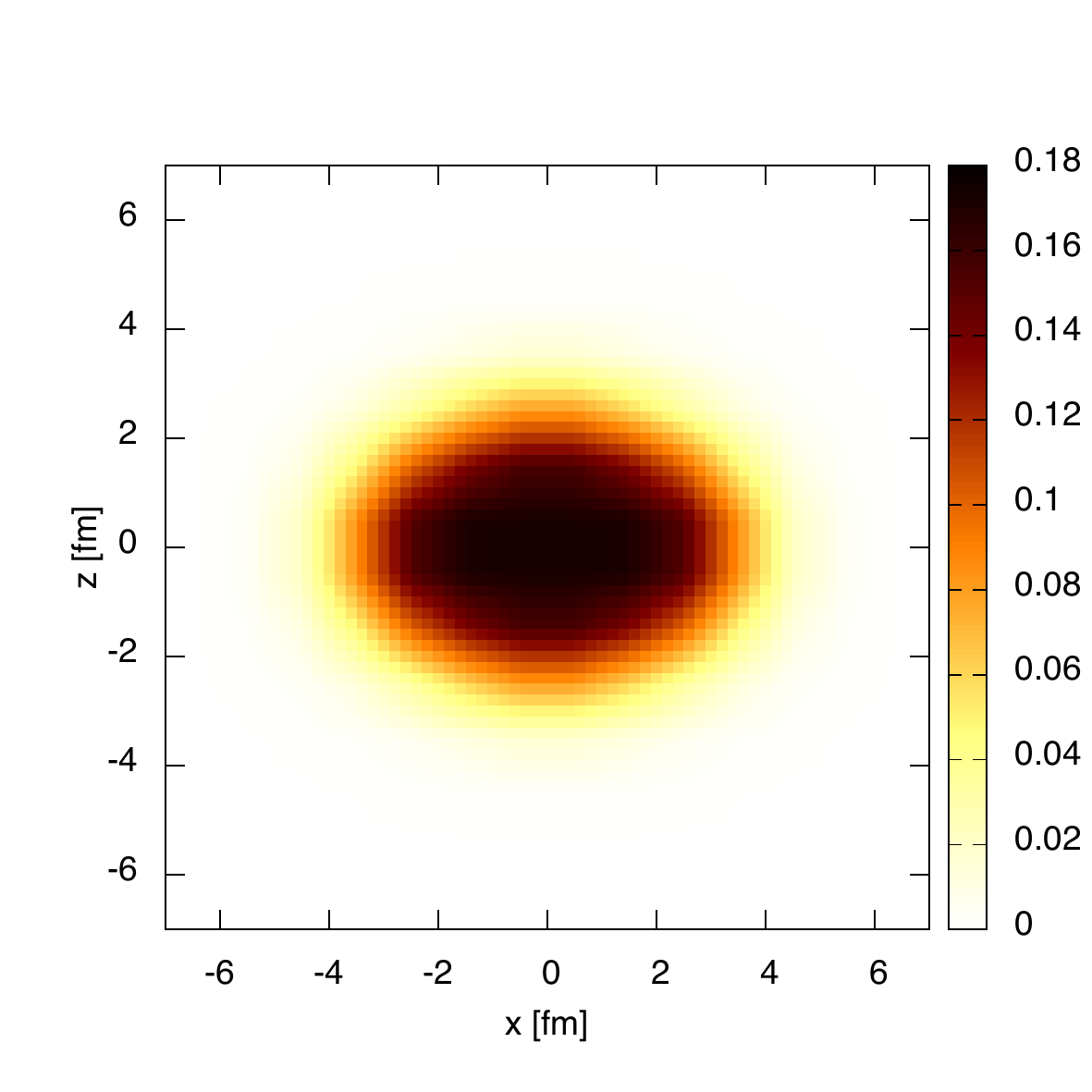}
  \caption{A slice of the calculated ground state density in $^{28}$Si using the SkI4 Skyrme interaction.  The nucleus is oblate and the total shape is given by envisioning the presented density slice rotated about the $x=0$ line.  The colour bar to the right of the figure gives the scale indicated in the total particle density presented in the main panel of the figure.\label{fig:si28den}}
\end{figure}

Giant resonances have long been known to show evidence of the underlying shape of the state upon which the resonances are built \cite{Dan64a,Dan64b}.  The underlying shape gives rise to a deformation splitting in which for a prolate nucleus, with two short axes and one long axis, there should be a higher peak at a higher frequency corresponding to vibrations in the short-axis directions, and a lower peak corresponding to the single mode in the long-axis direction.  In the case of an oblate nucleus, there are two long axes and one short axis, so the lower-energy modes dominate the higher-energy ones.  At least, this is the simple model.  In reality, there are other effects which confound a clear separation between deformation splitting, even within model calculations.  In time--dependent Hartree--Fock (or RPA, or other such models), Landau damping caused by the spread of underlying single particle energies can obscure the deformation splitting, in a way which can depend on details of the model \cite{Mar05}.

As an example here of the effect of shape on giant resonance vibrations, we perform a calculation of $^{28}$Si, which when calculated with the SkI4 Skyrme force \cite{Rei95} in the Sky3D code gave a static ground state deformation of $\beta_2=-0.210$.  A slice of the ground-state density in the $x$--$z$ plane is shown in figure \ref{fig:si28den}.

\begin{figure}[tb!]
  \includegraphics[width=\textwidth]{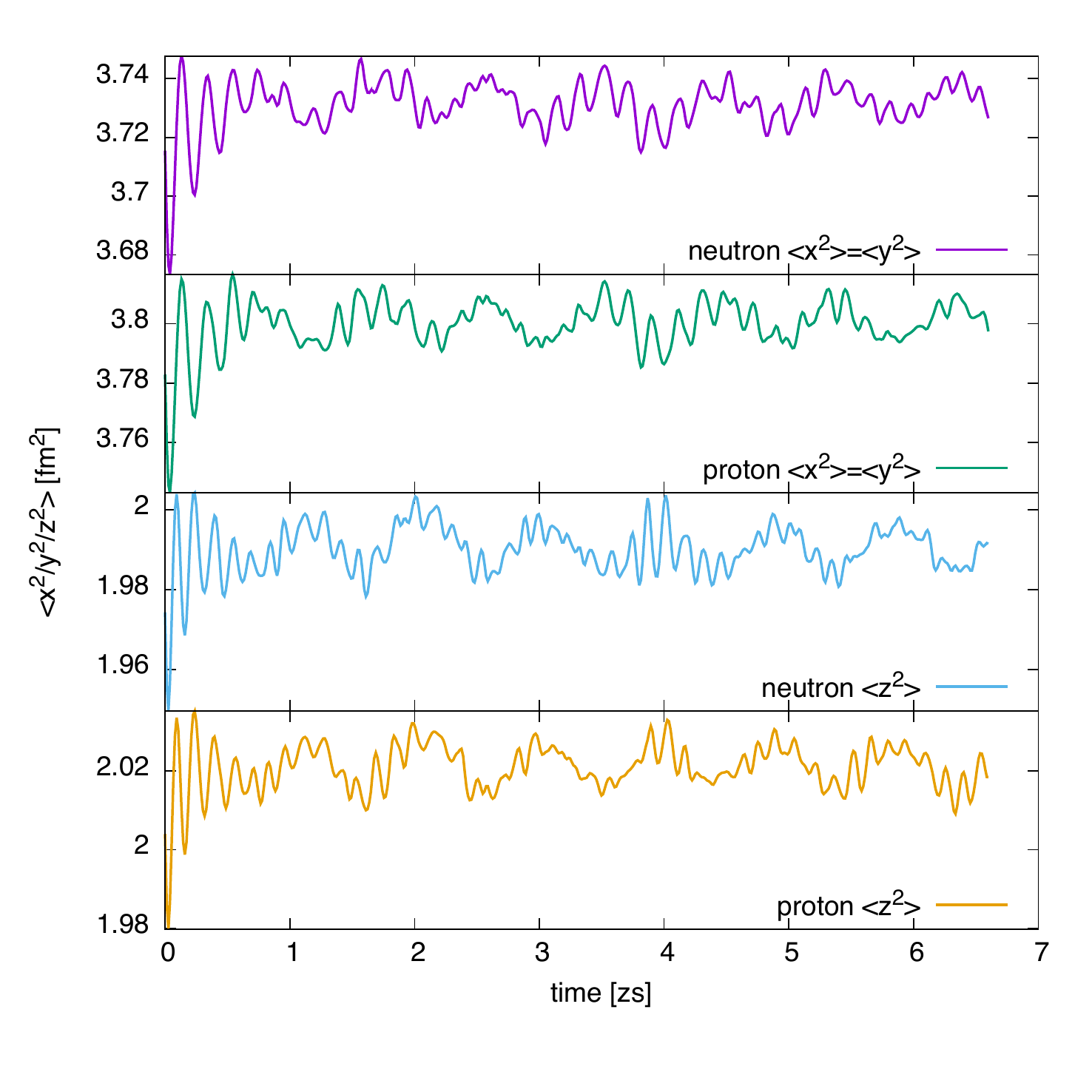}
  \caption{The time-dependent response of $^{28}$Si to an external monopole kick using the SkI4 Skyrme force.  The equal response in the two identical long axes $x$ and $y$ for neutrons and protons are shown in the top two panels, as labeled.  The response along the short axis, $z$, is shown for neutrons and protons in the final two panels.\label{fig:tresp}}
\end{figure}

We then proceed to give a monopole excitation to the nucleus in the form of an instantaneous boost of the form
\begin{equation}
  \psi(r,t=0^+)\rightarrow e^{ikr^2}\psi(r,t=0)
\end{equation}
where $\psi(t=0)$ are all the single--particle wave functions in the Slater Determinant, and $r$ is the coordinate relative to the centre of mass of the nucleus. $k$ is a parameter to control the strength of the imparted excitation.  The RPA limit is obtained as $k\rightarrow0$.  The calculations were performed with a version of the Sky3d code \cite{Mar14} to which modification were made in a particular way as a published exemplar for extending the code \cite{Ste14}.

The time--dependent response of the instantaneous excitation as a function of time is shown in Figure \ref{fig:tresp}.  Since the role of the $x$ and $y$ axes is symmetric for this nucleus, the response in the $x$ and $y$ direction is the same (the expectation value of $x^2$ for protons means $\int\rho_p(r)x^2d^3r$ etc.)  One sees by inspection of the responses along the $z$ direction (particularly by examining the period of the initial oscillations) that the dominant energy of excitation along the short axis is higher than along the two long axes, exactly as expected.

\begin{figure}[tb!]
  \includegraphics[width=\textwidth]{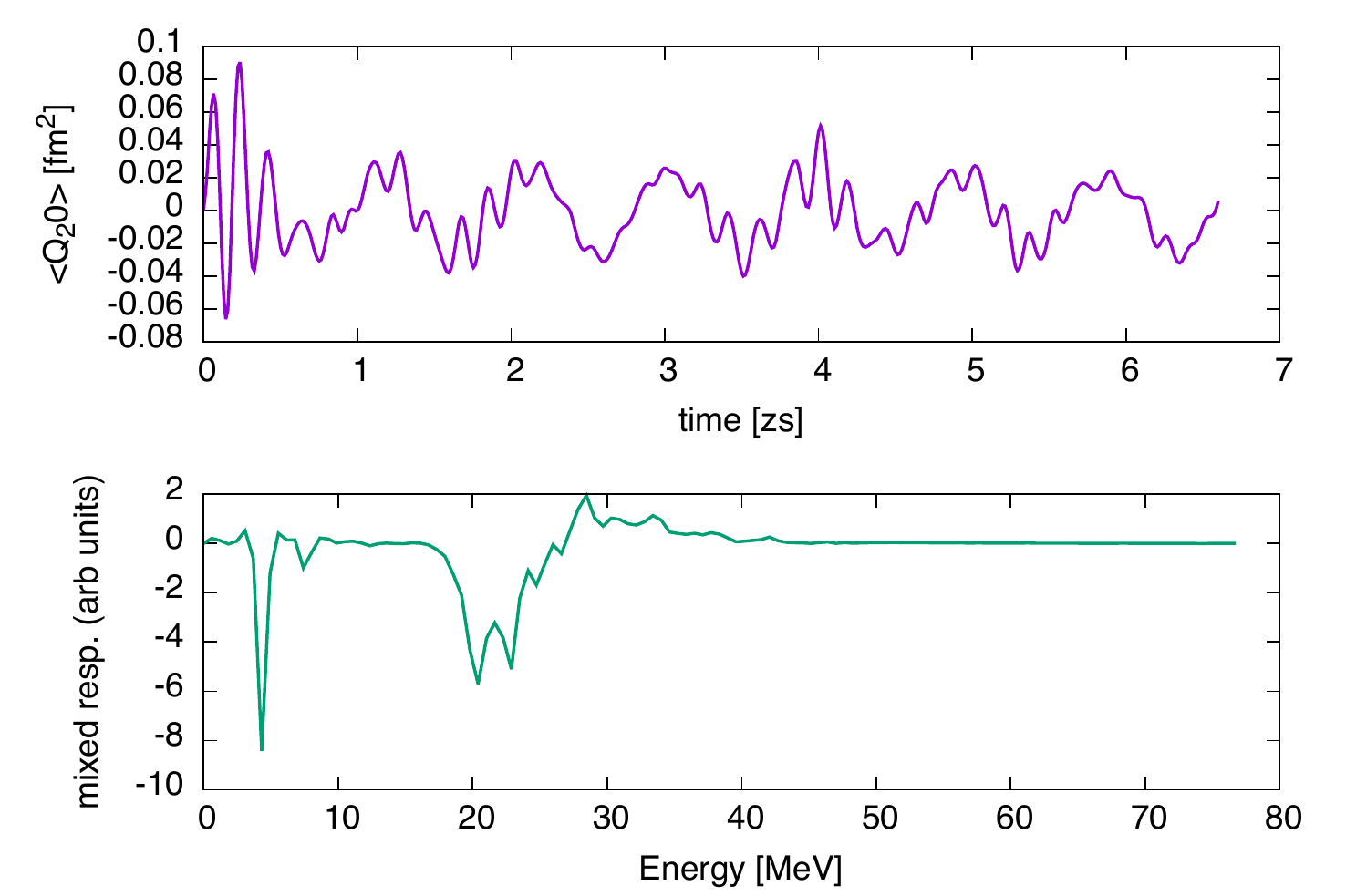}
  \caption{Quadrupole response to monopole kick (upper frame) and the imaginary part of its Fourier transform (lower frame). \label{fig:mixedresp}}
\end{figure}
 
One can also observe that there is a slow oscillation superimposed on the faster oscillations.  The interpretation of this is directly due to the deformation of the nucleus.  Because of the deformation, we see that the oscillations are not the same in all three directions, so therefore the initial ``monopole'' boost, immediately couples to a quadrupole excitation \cite{Sim03}, which must occur since $\langle Q_{20}\rangle=2\langle z^2\rangle-\langle x^2\rangle-\langle y^2\rangle\ne0$.  Figure \ref{fig:mixedresp} shows the $Q_{20}$ response of the $^{28}$Si nucleus to the monopole kick.  The top panel somewhat accentuates the quadrupole response and the lower panel shows the imaginary part of its Fourier Transform, which can be interpreted as a kind of mixed strength function as
\begin{equation}
  S(E) = \sum_\nu\langle0|\hat{M}|\nu\rangle\langle\nu|\hat{Q}|0\rangle\delta(E-E_\nu),
  \end{equation}
where $\hat{M}$ and $\hat{Q}$ denote the monopole and quadrupole operators respectively, $|0\rangle$ is the ground state and $|\nu\rangle$ the excitation state within the model space at energy $E_\nu$.  The non-zero nature of this strength function shows the coupling between the modes due to shape effects.  The sharp low energy peak in the strength is presumably the coupling to the first $2^+$ state.

\section{Shape fluctuations in Fission Isomers}

\begin{figure}[tb!]
  \includegraphics[width=\textwidth]{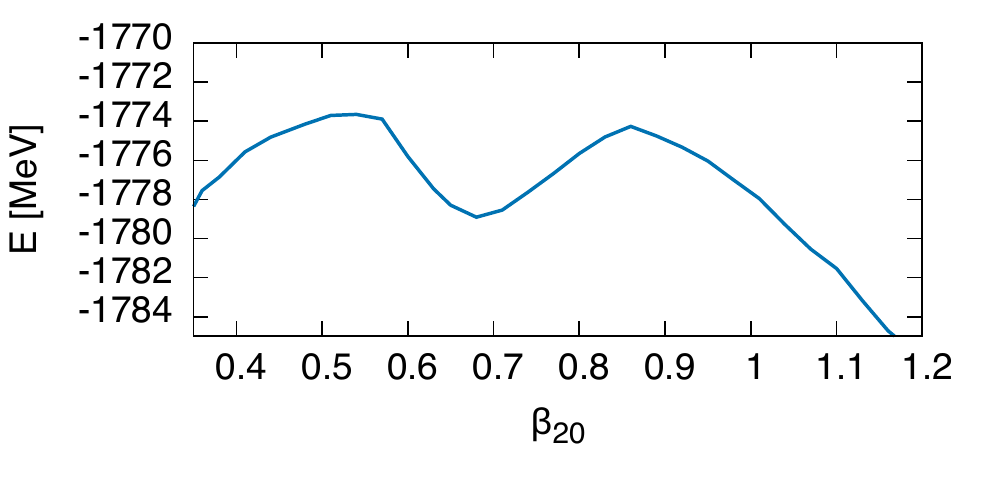}
  \caption{Potential energy surface around the fission isomer in $^{240}$Pu, calculated using the SkM* Skyrme force.\label{fig:pupes}}
\end{figure}

\begin{figure}[tb!]
  \includegraphics[width=\textwidth]{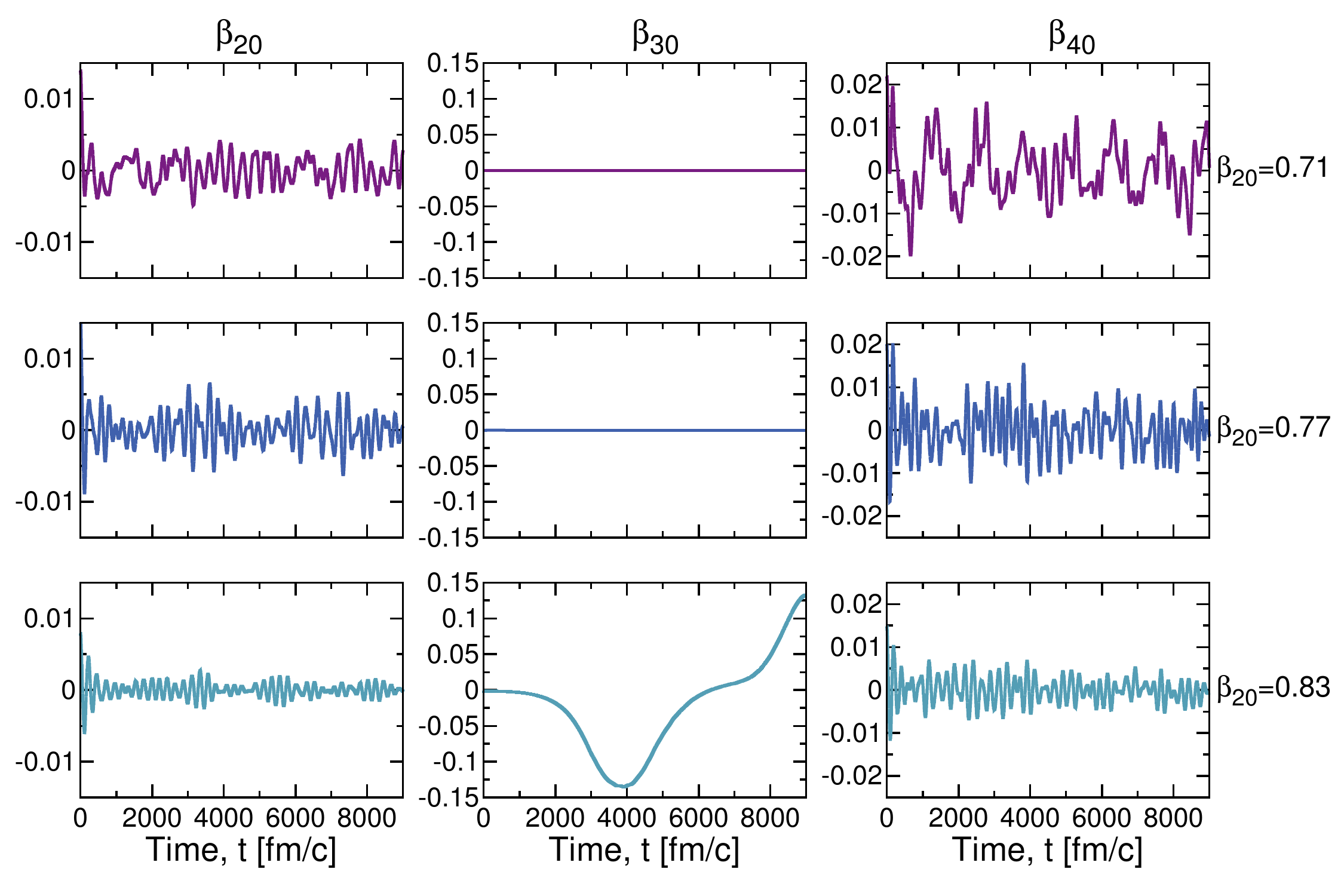}
  \caption{Time-evolution of different multipole deformation parameters as a function of time starting from different initial points along the constrained quadrupole deformation surface of $^{240}$Pu using the SkM*\cite{Bar82} force, which was adjusted to give good fission barriers.  The initial deformations are given at the right hand side of each row, and the deformation being followed is given as a column header.\label{fig:pu} }
  \end{figure}

As an example of starting a TDHF calculation from a state which is not an eigenstate of the static Hartree--Fock Hamiltonian, we performed calculations of the time-dependent response following the release of a static configuration which had been constrained to have a particular quadrupole deformation, inside the fission isomer valley in $^{240}$Pu.  These calculations aimed to understand general non-adiabatic fission properties \cite{God14,God15a,God15b}.  The potential energy surface around the fission isomer is seen in Figure \ref{fig:pupes}.  It shows the fission isomer with a deformation of around $\beta_{20}\simeq0.69$ with a barrier height of around 5 MeV.  The ground state is off to the left of this zoomed-in plot.

We performed calculations starting at different points on the slope of the fission isomer valley, leading towards the fission barrier, at deformations of $\beta_{20}=0.71$, $\beta_{20}=0.77$, and $\beta_{20}=0.83$.  Following the subsequent shape vibrations as a function of time, a range of interesting behaviours were evident, and are shown in Figure \ref{fig:pu}.  One can see qualitatively different results between the three different release points, especially in the way that the most quadrupole-deformed initial shape manages to explore a region of octupole deformation, while others do not.  We see also hexadecupole fluctuations, but did not take the studies further in terms of shape analysis as it was not the focus of this fission study.

We made quite extensive studies of different release points also after the barrier, including fissioning states, where the TDHF method proved amenable to following shape properties also of fission fragments.  We direct the interested reader to the resulting publications \cite{God14,God15a,God15b}.

\section{Conclusion}
Ways in which nuclear shape influence dynamics (at least from a TDHF perspective) have been presented.  The link between giant resonances on deformed nuclei and the resulting different response in along different intrinsic axes in the nucleus (which is ultimately responsible for deformation splitting) was shown, along with a consequence for coupling of different shape excitations.  We gave also a recent indicative calculations of shape vibrations in the fission isomer of $^{240}$Pu showing that multiple multipoles come into play.  We conclude that it is clear that TDHF provides a useful tool for understanding nuclear shapes.

\section*{Acknowledgments}
Support for this work comes from the UK STFC funding council via grants ST/L005743/1 and ST/J000051/1, along with the award of time on the DiRAC computer system, and via the Bulgarian National Science Fund (BNSF) under Contract No. DFNI-E02/6.

\end{document}